\documentclass{article}
\usepackage{booktabs}
\usepackage{amsmath}
\usepackage{amssymb}
\usepackage{bm}
\usepackage{etoolbox}
\usepackage{array,ragged2e}
\usepackage{tikz}
\usepackage{csquotes}
\usetikzlibrary{decorations.pathreplacing}
\usetikzlibrary{automata,positioning}
\usepackage{graphicx}
\graphicspath{{./}}
\usepackage{array}
\usepackage{xspace}

\usepackage[backend=biber,style= numeric-comp,sorting=none,maxcitenames=2, maxbibnames=99]{biblatex}
\renewbibmacro{in:}{%
	\ifentrytype{article}{}{\printtext{\bibstring{in}\intitlepunct}}}
\usepackage{environ}
\usepackage[justification=centering]{caption}
\usepackage{subcaption}
\usepackage{slashbox}
\usepackage{longtable}
\usepackage{mwe}

\DeclareMathOperator*{\E}{\mathbb{E}}
\newcolumntype{H}{>{\setbox0=\hbox\bgroup}c<{\egroup}@{}}

\newcounter{midruleV}
\newcommand*{\midruleV}{%
	\aftergroup\aftergroup\aftergroup\midruleVaux
}
\newif\ifmidruleV
\makeatletter
\newcommand*{\midruleVaux}{%
	\noalign{%
		\stepcounter{midruleV}%
		\ifnum\value{midruleV}=5 %
		\global\midruleVtrue
		\setcounter{midruleV}{0}%
		\else
		\global\midruleVfalse
		\fi
	}
	\ifmidruleV\midrule\fi
}
\newcommand*{\resetmidruleV}{\setcounter{midruleV}{0}}
\makeatother
\date{}

\begin{document}

\title{Comparing Alternatives to Measure the Impact of DDoS Attack Announcements on Target Stock Prices.}

\maketitle

\author{\renewcommand{\arraystretch}{0.8}
	\begin{tabular}{ c c c}
		\small Abhishta & \small Reinoud Joosten & \small L.J.M. Nieuwenhuis\\
		\footnotesize University of Twente & \footnotesize University of Twente & \footnotesize University of Twente\\
		\footnotesize The Netherlands & \footnotesize The Netherlands & \footnotesize The Netherlands\\
		\footnotesize s.abhishta@utwente.nl & \footnotesize r.a.m.g.joosten@utwente.nl  & \footnotesize l.j.m.nieuwenhuis@utwente.nl			
	\end{tabular}\\
}

%------------------------------------------------------------------------------
% Abstract
%
\begin{abstract}
	The attack intensity of distributed denial of service (DDoS) attacks is increasing every year. Botnets based on internet of things (IOT) devices are now being used to conduct DDoS attacks. The estimation of direct and indirect economic damages caused by these attacks is a complex problem. One of the indirect damage of a DDoS attack can be on the market value of the victim firm. In this article we analyze the impact of 45 different DDoS attack announcements on victim's stock prices. We find that previous studies have a mixed conclusion on the impact of DDoS attack announcements on the victim's stock price. Hence, in this article we evaluate this impact using three different approaches and compare the results. In the first approach, we use the assume the cumulative abnormal returns to be normally distributed and test the hypothesis that a DDoS attack announcement has no impact on the victim's stock price. In the latter two methods, we do not assume a distribution and use the empirical distribution of cumulative abnormal returns to test the hypothesis. We find that the assumption of cumulative abnormal returns being normally distributed leads to overestimation/underestimation of the impact. Finally, we analyze the impact of DDoS attack announcement on victim's stock price in each of the 45 cases and present our results.    
\end{abstract}

%------------------------------------------------------------------------------
\section{Introduction and Background}
\label{Introduction}

\footnotetext[1]{The final version of this paper has been published in Journal of Wireless Mobile Networks, Ubiquitous Computing, and Dependable Applications (JoWUA), Volume 8, Number 4.}

DDoS attacks are responsible for creating unavailability of online resources which can lead to both direct and indirect losses \cite{Anderson2013}. In 2016 the intensity of DDoS attacks peaked at 1.4 Tb/s. The biggest distributed denial of service attack targeted the systems operated by DNS provider Dyn \cite{dyn}. A few months later this firm was bought by Oracle \cite{oracle}. One can only speculate about the change in the valuation of the firm as it is not publicly traded. In this study we investigate the impact of DDoS attack announcements on the stock price of the victim firms.

\begin{figure}[h!]
	\centering
	\includegraphics[width=\textwidth]{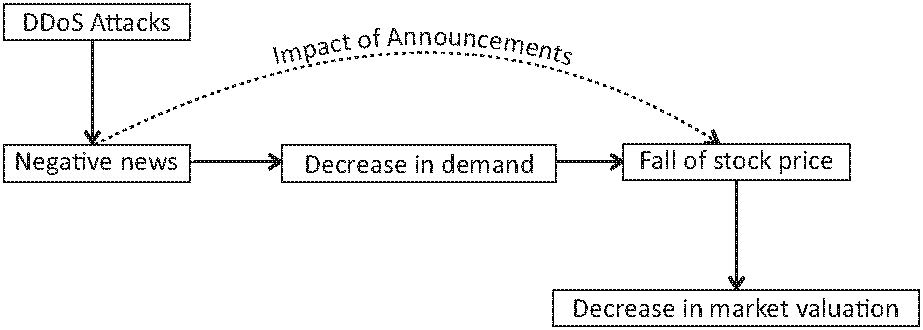}
	\caption{Impact of a DDoS attack announcement on market valuation of the firm }
	\label{fig:ddos to market}
\end{figure} 	

The stock price of a firm is representative of its market value. In the past economists have analyzed the impact of an economic event on the value of the firm \cite{Mackinlay1997}. A strategic business decision e.g. \textit{merger or an acquisition} can significantly impact the future dividends. For instance, in the case of a possible negative impact on the future cash flows, it is beneficial for the investors to sell the shares and invest in a different stock.

DDoS attacks may lead to negative news articles about the firm. These news articles come as a negative sentiment shock and can negatively influence the demand of the victim firm's shares \cite{tetlock2007}. This in-turn leads to the fall of stock prices of the attacked company. Figure \ref{fig:ddos to market} shows the conceptual relationship between DDoS attack events and decrease in market valuation of the victim firm. It also shows the empirical link that we investigate in this article.
      
Estimating the impact of cyber security related events is a complex problem \cite{Florencio2013,Santos1993}. Several studies have tried to investigate the impact of cyber security related announcements on the victim stock prices and we discuss the results and limitations of these studies in Section \ref{Previous Works}. In this article we use three different methods for analyzing the impact of these attack announcements on target stock prices and then discuss and explain the differences in results.

This is an extended version of our study \cite{Joosten2017} that analyzed the impact of DDoS attack announcements on victim stock prices. In this article we compare the method proposed in \cite{Joosten2017} with the traditional methods of event study and illustrate the disadvantages of using the assumptions and approximations considered in those. We also analyze an extended set of DDoS attack announcements and re-emphasize the results of our previous study.

\section{Related Literature}
\label{Previous Works}

Event studies have been used by researchers to study the impact of various firm related announcements on the stock price. \citeauthor{Mackinlay1997} \cite{Mackinlay1997} discussed a method of conducting an event study including the various market estimation models. In this section we discuss the articles that have contributed to evaluation of the impact cyber security event announcements have on victim stock prices.

\citeauthor{HovavAnatandDArcy2003} \cite{HovavAnatandDArcy2003} used a so-called one-factor market model in order to estimate the stock prices. Equation \ref{market model} shows the estimation model used by them, where $r_{it}$ represents the return rate of the stock $i$ on day $t$ and $r_{mt}$ represents the rate of return of the market index on day $t$. As an example, $r_{it}$ can be computed as $(P_{it}-P_{it-1})/P_{it-1}$, where $P_{it}$ is the price of the stock on day $t$. The parameters $\alpha_{i}$ and $\beta_{i}$ are firm dependent coefficients and can be estimated using ordinary least square (OLS). The stochastic variable $\epsilon_{it}$ is the error term with $\E{[\epsilon_{it}]}=0$. \citeauthor{HovavAnatandDArcy2003} \cite{HovavAnatandDArcy2003} analyzed a sample of 23 announcements of denial of service attacks and were not able to find any significant impact of these announcements on the capital market.   

\begin{equation}
\label{market model}
r_{it}={\alpha_i}+{\beta_i}r_{mt} + \epsilon_{it}
\end{equation}

Later, \citeauthor{Campbell2003} \cite{Campbell2003} used the above discussed estimation model to analyze a sample of 43 announcements of all kinds of cyber attacks. The abnormal returns were calculated by them using Equation \ref{addar} and the cumulative abnormal returns ($CAR$) were computed with the use of Equation \ref{addcar}. They assumed these $CAR$s to be normally distributed and used a $Z$-statistic to test their hypothesis (i.e. there was no impact of cyber attack announcements on victim stock prices) and reported significant negative impact due to information security breach announcements.

\begin{equation}
\label{addar}
AR_{it}=r_{it}-(\hat{\alpha_i}+\hat{\beta_i}r_{mt})
\end{equation}

\begin{equation}
\label{addcar}
CAR_n=\sum_{t=-1}^{n}AR_{it}
\end{equation}

\citeauthor{Cavusoglu2004} \cite{Cavusoglu2004} and \citeauthor{Kannan2007} \cite{Kannan2007} also use the above described method for analyzing the impact of security breach announcements. The former concluded that these announcements not only influence the value of the announcing firms but also the value of their internet security developers. While the later considered a sample of 102 and reported a decrease of 1.4\% in the market valuation relative to the control group.

\citeauthor{Gordon2011} \cite{Gordon2011} used a so-called three factor Fama-French model \cite{famafrench} for the estimation. This model estimates the stock price on the basis of company size, company price-to-book ratio, and market risk, and can be mathematically represented as shown in Equation \ref{famafrench}. $SMB_t$ is the difference between the return on the portfolio of small stocks and the return on the portfolio of large stocks on day $t$, and $HML_t$  is the difference between the return on a portfolio of low-book-to-market stocks and the return on a portfolio of low-book-to-market stocks on day $t$. The parameters ${a_i}$,${b_i}$,${s_i}$ and ${h_i}$ are Fama and French three-factor model estimated firm-dependent coefficients. The stochastic variable $\epsilon_{it}$ is the error term with $\E{[\epsilon_{it}]}=0$. \cite{Gordon2011} reported no significant impact due to post 9/11 announcements.

\begin{equation}
\label{famafrench}
r_{it}={a_i}+{b_i}r_{mt}+{s_i}SMB_t+{h_i}HML_t+\epsilon_{it}\\
\end{equation}   

These mixed results motivate us to evaluate the impact of the choice of model and the underlying assumptions in the study on the final results. Thus, in this article we evaluate the impact of DDoS attack announcements on victim stock prices using three different methods and compare their results in Section \ref{Results}. Section \ref{Method} discusses the methodology used by us in this study.  

\section{Methodology}
\label{Method}

\begin{figure}[ht!]
	\centering
	\includegraphics[width=\textwidth]{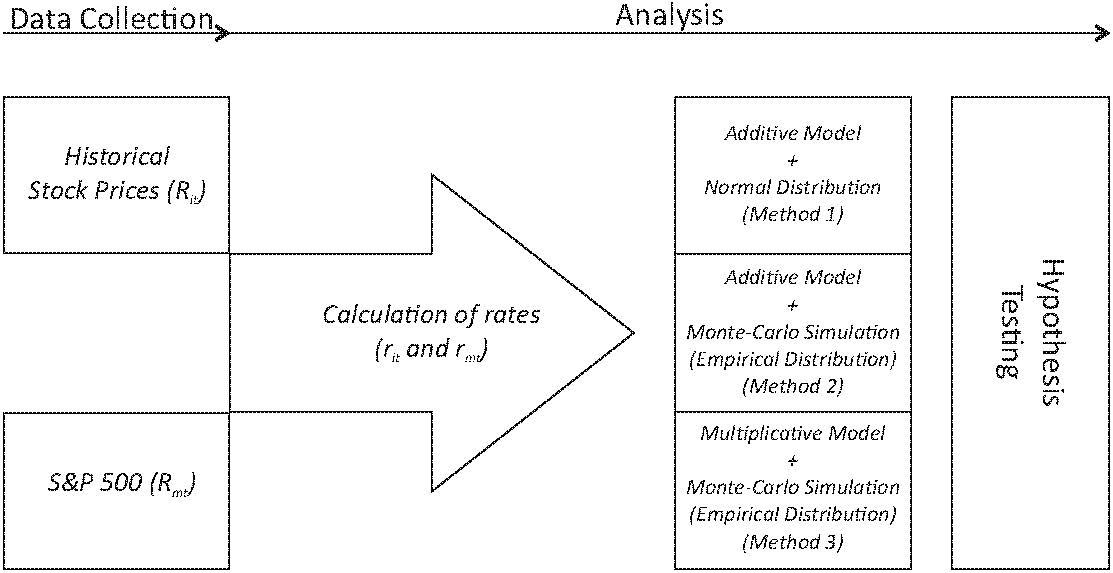}
	\caption{Methodology for this study.}
	\label{Method Diagram}
\end{figure} 	

The methodology used by us can be broadly subdivided in two parts:
\begin{enumerate}
	\item Data Collection
	\item Analysis
\end{enumerate}
In this study we analyze the impact on stock returns using three different methods. Firstly, we use the event study method employed by many of the previous articles \cite{Campbell2003,HovavAnatandDArcy2003,Cavusoglu2004}. In the second method, we use an additive market model for the estimation of return rates and then use the empirical distribution of abnormal returns by generating random scenarios for analyzing the additive cumulative abnormal return. In the last method we use the method proposed by us, that makes use of a multiplicative model for estimation and later uses multiplicative cumulative abnormal returns for analysis \cite{Joosten2017}. Figure \ref{Method Diagram} illustrates the step by step process used.

\subsection{Data Collection}
\label{Data Collection}

The data set in this study consists of all DDoS attack announcements made on the web since December, 2010. The final list of announcements that were evaluated for this study are shown in Table \ref{data and summary}. It also shows the total number of negative, positive and no impact periods in each case. In total 60 DDoS attack announcements were considered for this study. We further filter these announcements on the basis of the following criteria:

\begin{itemize}
	\setlength\itemsep{0em}
	\item In case of multiple announcements made on consecutive days, the earliest announcement was considered.
	\item All announcements in relation with companies that were not publicly traded at the time attack were removed from the dataset.
	\item All such announcements that reported DDoS attacks were coupled with integrity and confidentiality attacks were not considered. This was done to analyze the impact of DDoS attack announcements in isolation on the company's stock price.    
\end{itemize}

The above criterion of filtering is consistent with previous studies \cite{Joosten2017}. Yahoo! finance was used in order to collect stock prices for all the firms. We use S\&P 500 index values for calculating the market rate ($r_{mt}$). Standard and Poor's (S\&P) 500 has been used by many of the previous studies as the index of the market. Finally, after filtering the initial dataset we analyze a sample of 45 announcements. 

\subsection{Analysis}
\label{Analysis}
For analysis of the data set we first establish the null hypothesis ($H_0$) as follows:

\begin{description}
	\item[$H_0$:] \emph{There is no impact of DDoS attack announcements on victim stock prices.}
\end{description}

In order to analyze the collected data we first need to calculate the rate of return of the market index on day $t$ ($r_{mt}$) and $r_{it}$ the rate of return of the stock $i$ on day $t$. The rate of return can be calculated as shown in Equation \ref{rate}, where $R_{it}$ and $R_{mt}$ represent the stock price and market index for day $t$. The value of the market index shows the average of returns of all the firms included in the market index. 

\begin{equation}
\label{rate}
\begin{split}
r_{it} &= \dfrac{R_{it}-R_{i(t-1)}}{R_{i(t-1)}}\\
r_{mt} &= \dfrac{R_{mt}-R_{m(t-1)}}{R_{m(t-1)}}\\
\end{split}
\end{equation}

In this study we use three different methods to test our null hypothesis ($H_0$). After explaining in detail these methods in Sections \ref{method 1}, \ref{method 2} and \ref{method 3} we then compare the results in Section \ref{Results} and conclude in Section \ref{conclusion}.

\subsubsection{Method 1}
\label{method 1}

In the first method we consider an additive model to represent the normal behavior of the market. The model can be mathematically represented as shown in Equation \ref{add_model}. This model is used to estimate the returns on a firm's stock. The parameters $r_{it}$ and $r_{mt}$ are calculated as shown in Equation \ref{rate}.

\begin{equation}
\label{add_model}
r_{it} = \alpha_i+ \beta_i{r_{mt}}+\epsilon_{it}\\
\end{equation}

The stochastic variable $\epsilon_{it}$ is the error term with $\E{[\epsilon_{it}]}=0$. We use ordinary least square (OLS) in order to calculate the estimations $\hat{\alpha_i}$ and $\hat {\beta_i}$ for the firm dependent parameters $\alpha_{i}$ and $\beta_{i}$ by considering daily returns over a period of 200 days. The estimation period starts 201 days before the date of attack announcement and ends two days before the announcement.

\begin{figure}[h]
	\begin{center}
		\begin{tikzpicture}
		\draw[->,ultra thick] (-5,0)--(1.5,0);
		\draw[-,thick] (-4.8,0)node[below]{\small{$-201$}}--(-4.8,2) ;
		\draw[-,thick] (-1,0)node[below]{\small{$-1$}}--(-1,3) ;
		\draw[-,thick] (-1.40,0)node[below]{\small{$-2$}}--(-1.40,2) ;
		\draw[-,thick] (0.5,0)node[below]{\small{$7$}}--(0.5,2.25) ;
		\draw[-,thick] (1,0)node[below]{\small{$9$}}--(1,3) ;
		\draw[-,thick] (-0.6,0)node[below]{\small{$1$}}--(-0.6,0.75);
		\draw[-,thick] (-0.2,0)node[below]{\small{$3$}}--(-0.2,1.25);
		\draw[-,thick] (0.2,0)node[below]{\small{$5$}}--(0.2,1.75);
		\draw[<->] (-4.8,2)--(-1.40,2) node[midway,below]{\tiny{Estimation Period}};
		\draw[<->] (-4.8,2)--(-1.40,2) node[midway,above]{\tiny{$[-201,-2]$}};
		\draw[<->] (-1,3)--(1,3) node[midway,below]{\tiny{$[-1,9]$}};
		\draw[<->] (-1,2.25)--(0.5,2.25) node[midway,above]{\tiny{$[-1,7]$}};
		\draw[<->] (-1,0.75)--(-0.6,0.75) node[right,above]{\tiny{$[-1,1]$}};
		\draw[<->] (-1,1.25)--(-0.2,1.25) node[midway,above]{\tiny{$[-1,3]$}};
		\draw[<->] (-1,1.75)--(0.2,1.75) node[midway,above]{\tiny{$[-1,5]$}};
		\draw[<->] (-1,3)--(1,3) node[midway,above]{\tiny{Event Periods}};
		\end{tikzpicture}
	\end{center}
	\caption{Estimation and Event Periods.}
	\label{periods}
\end{figure}
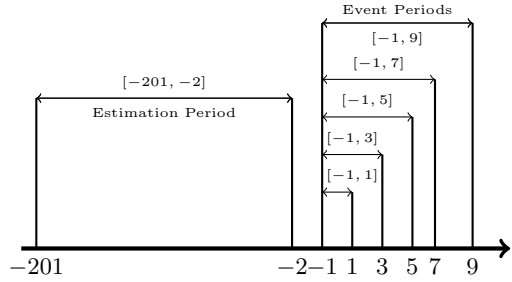    

The additive abnormal return ($AAR_{it}$) is the measurement of the deviation of the actual returns from the ones calculated with the help of additive model equation \ref{add_model}. Hence $AAR_{it}$ can be mathematically represented as:

\begin{equation}
\label{eq:AAR}
AAR_{it} = r_{it}-(\hat{\alpha_i}+\hat{\beta_i}r_{mt})\\
\end{equation}

We measure the impact of DDoS attack announcements on the stock return over the following five \emph{event periods}:

\begin{enumerate}
	\item One day prior to the announcement to 1 days after it $[t-1,t+1]$.
	\item One day prior to the announcement to 3 days after it $[t-1,t+3]$.
	\item One day prior to the announcement to 5 days after it $[t-1,t+5]$.
	\item One day prior to the announcement to 7 days after it $[t-1,t+7]$.
	\item One day prior to the announcement to 9 days after it $[t-1,t+9]$.
\end{enumerate}

We keep these time periods consistent for all methods. The \emph{estimation period} and the \emph{event periods} are shown in Figure \ref{periods}. We take the event periods from one day prior to the announcements in order to compensate for any time lags. In order to calculate the combined effect over a certain number of days, we calculate the additive cumulative abnormal return ($ACAR$) as shown in Equation \ref{ACAR} for the period $[N_1,N_2]$.

\begin{equation}
\label{ACAR}
ACAR_i= \sum_{t=N_1}^{N_2}(AAR_{it})
\end{equation}

We compute the mean $ACAR$ for 45 events in our sample as follows:

\begin{equation}
\label{mean_ACAR}
ACAR = \dfrac{1}{K}\sum_{i=1}^{K}ACAR_i\\
\end{equation} 

Where $K$ is the number of events. We then estimate the standard deviation ($\sigma_{ACAR}$) using Equation \ref{SD_ACAR}. 

\begin{equation}
\label{SD_ACAR}
\sigma_{ACAR}=\sqrt{\dfrac{\sum_{i=1}^{K}(ACAR_i-ACAR)^2}{K-1}}\\
\end{equation}

\begin{figure}[h]
	\centering
	\includegraphics[width=\textwidth]{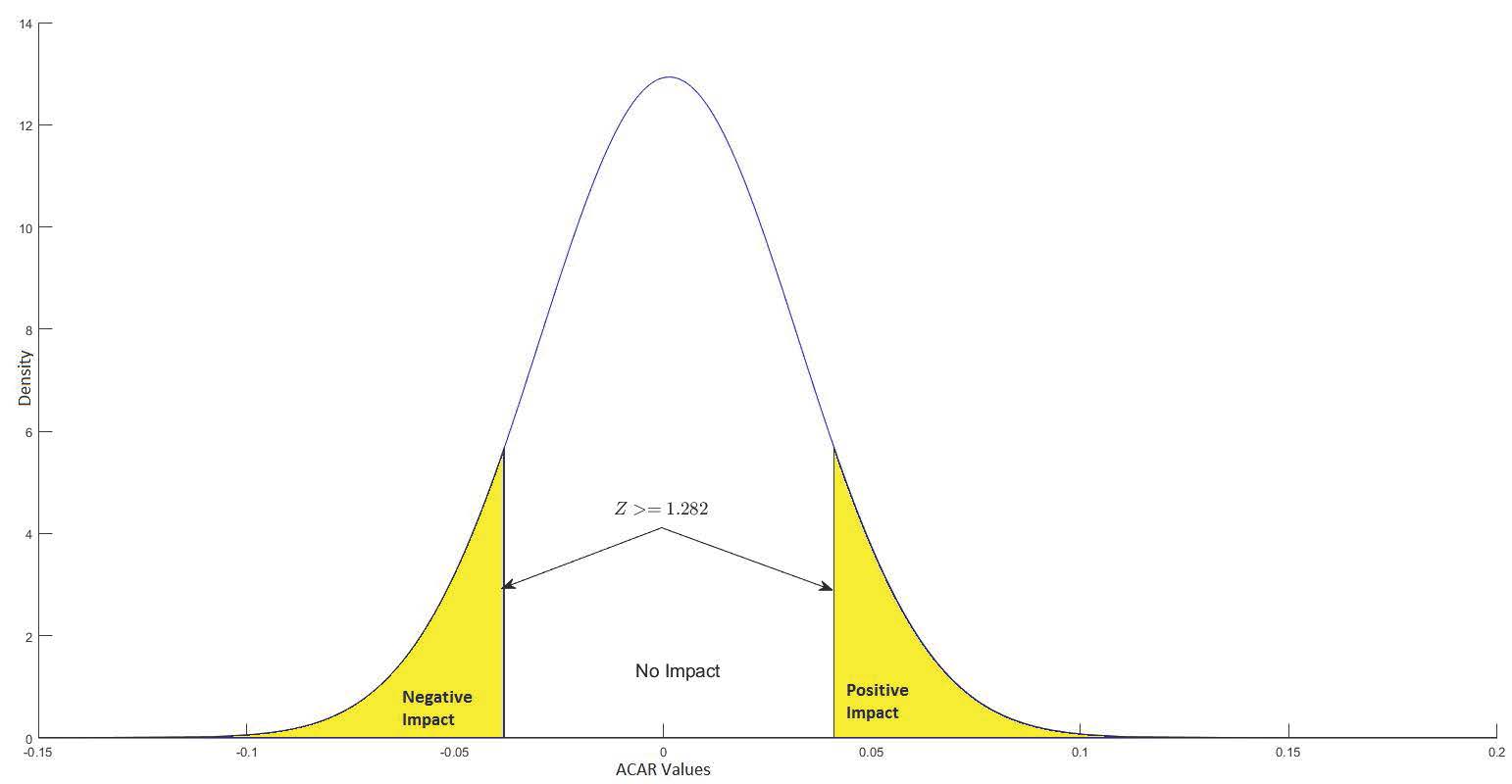}
	\caption{Normal distribution for 5 day $ACAR$ values and decision rule for impact analysis.}
	\label{fig:normal distribution}
\end{figure}

We now assume the $ACAR_i$ values for a given \emph{event period} to be normally distributed and test for significance by making use of the $Z$-statistic at 10\% confidence level. Hence we reject the null hypothesis if the $|Z|>=1.282$ as shown in Figure \ref{fig:normal distribution}.

\subsubsection{Method 2}
\label{method 2}

In this method we again make use of the additive estimation model as shown in Equation \ref{add_model}. We avoid the widespread assumption of short-term returns being approximately normally distributed. We also do not impose any alternative distribution to these returns. Instead we use the technique of bootstrapping (e.g. \citeauthor{efron1992bootstrap} \cite{efron1992bootstrap}). In this case we generate 5 million $n$-day returns by randomly drawing $n$ one-day returns from the empirical distribution. The relative frequencies of these 5 million multi-day returns are then used as the distribution for hypothesis testing.

\begin{figure*}[t]
	\centering
	\begin{subfigure}[b]{0.475\textwidth}
		\centering
		\includegraphics[width=\textwidth]{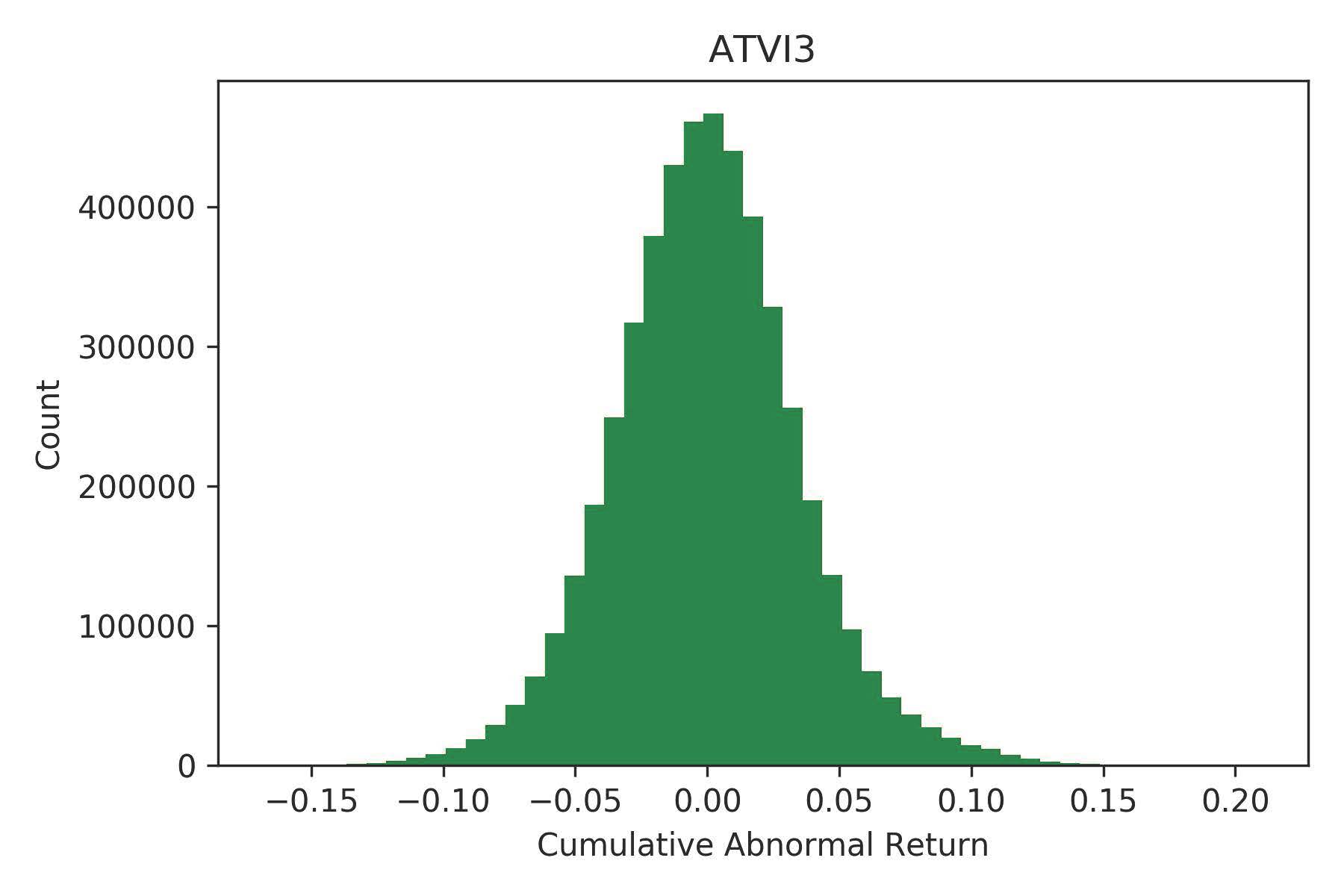}
		\caption[]%
		{{\small 3-Day $ACAR_{Activision Blizzard}$}}    
		\label{3-day add}
	\end{subfigure}
	\hfill
	\begin{subfigure}[b]{0.475\textwidth}  
		\centering 
		\includegraphics[width=\textwidth]{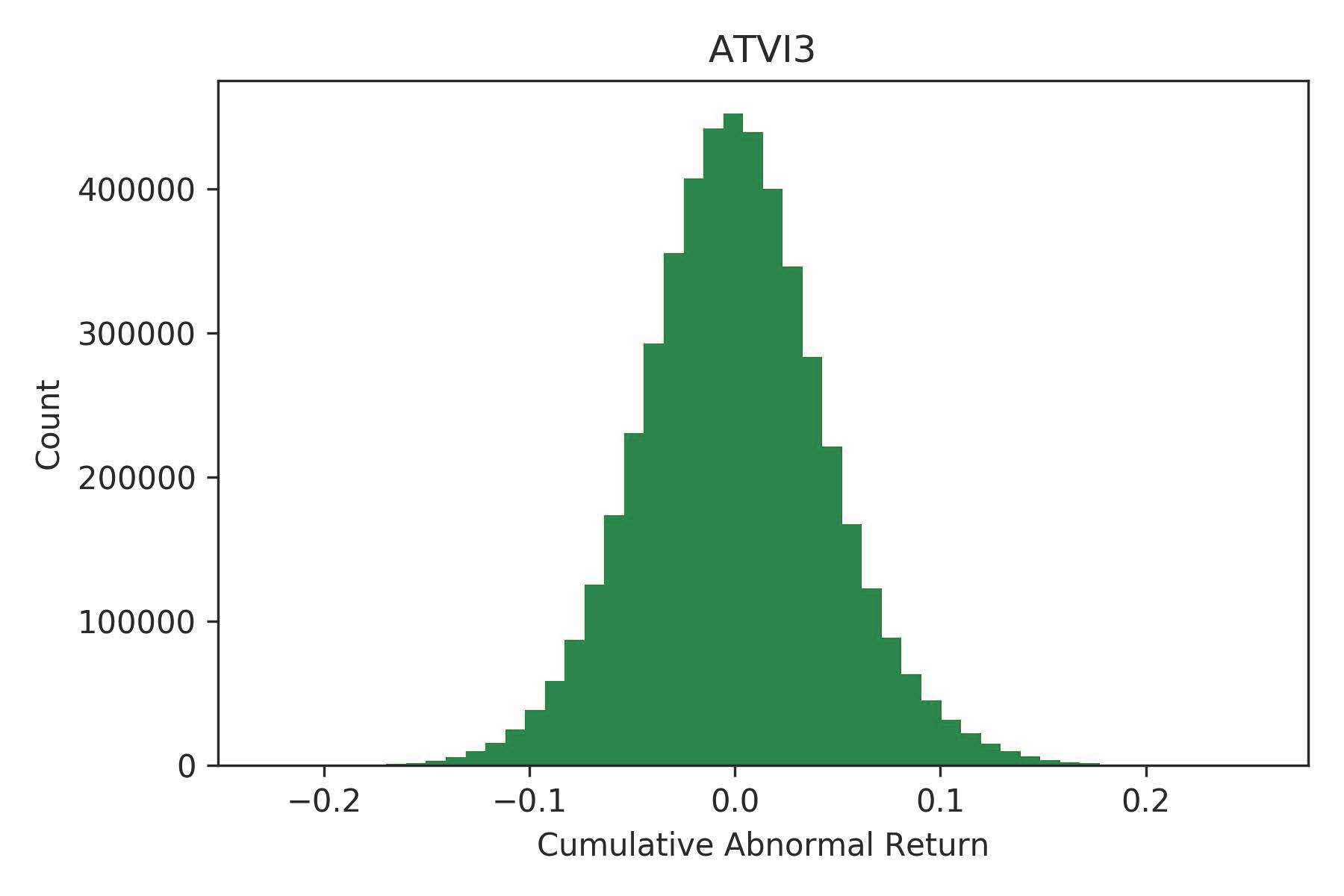}
		\caption[]%
		{{\small 5-Day $ACAR_{Activision Blizzard}$}}    
		\label{5-day add}
	\end{subfigure}
	\vskip\baselineskip
	\begin{subfigure}[b]{0.475\textwidth}   
		\centering 
		\includegraphics[width=\textwidth]{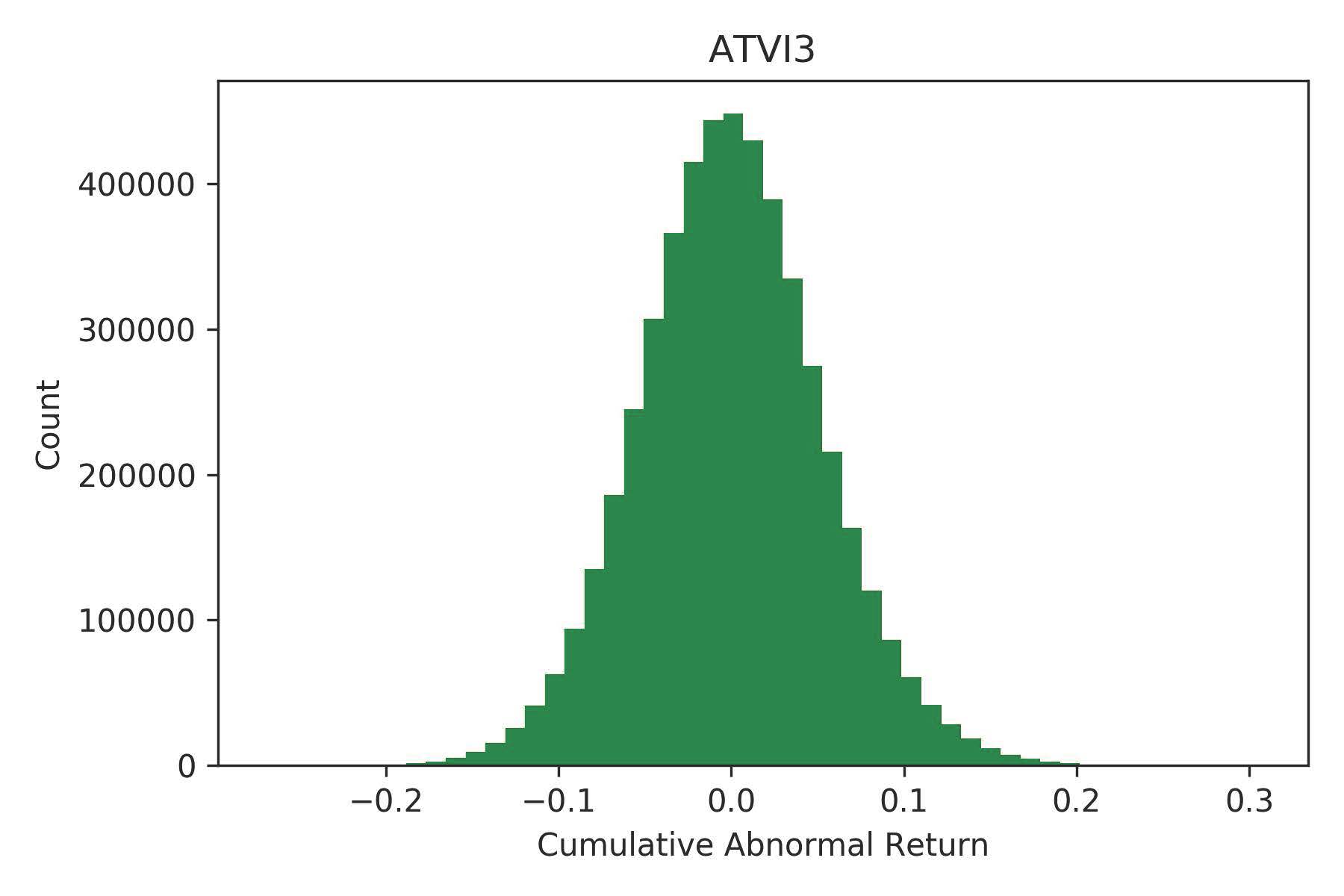}
		\caption[]%
		{{\small 7-Day $ACAR_{Activision Blizzard}$}}    
		\label{7-day add}
	\end{subfigure}
	\quad
	\begin{subfigure}[b]{0.475\textwidth}   
		\centering 
		\includegraphics[width=\textwidth]{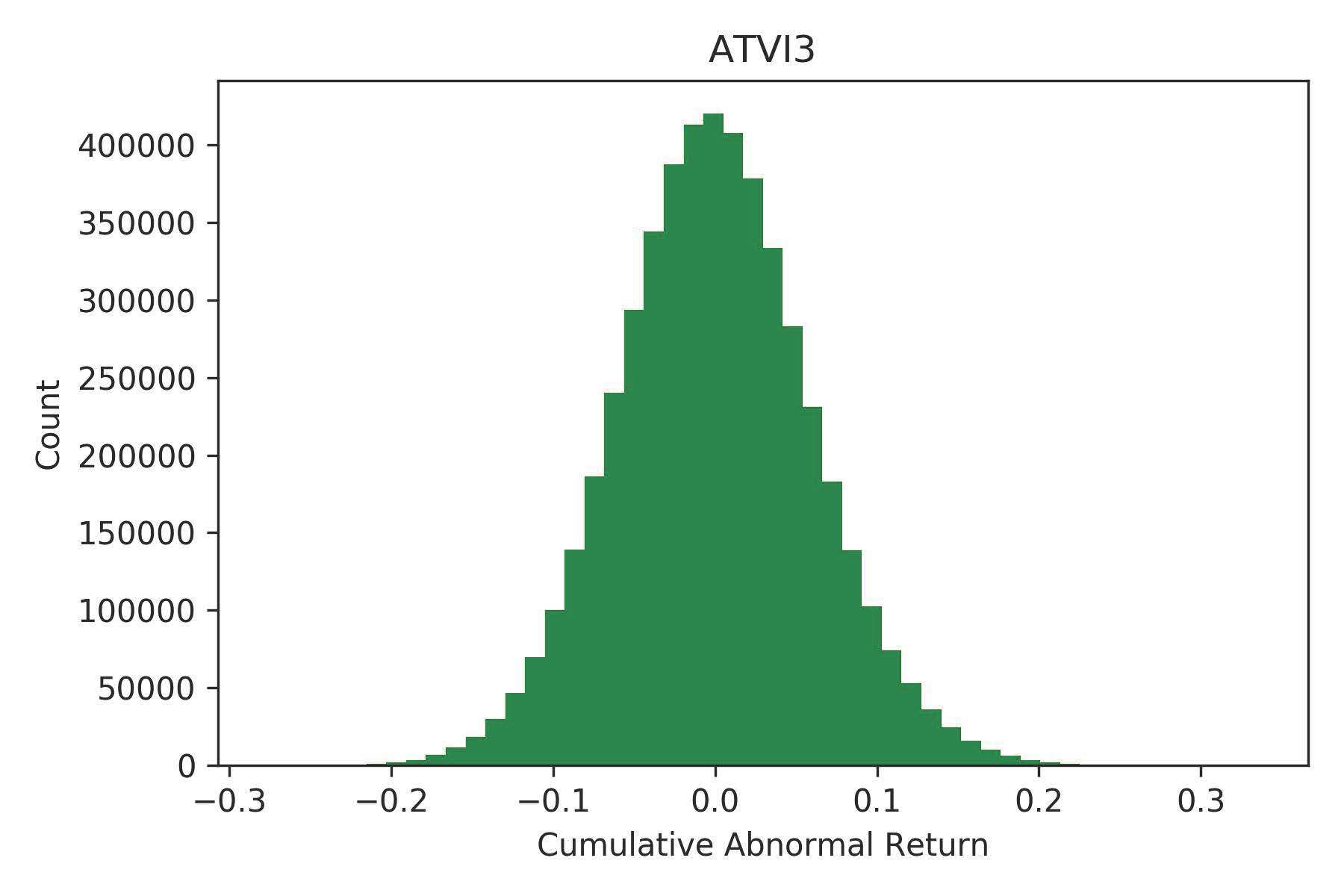}
		\caption[]%
		{{\small 9-Day $ACAR_{Activision Blizzard}$}}    
		\label{9-day add}
	\end{subfigure}
	\caption[]
	{\small Empirical distribution of $ACAR$(additive) for Activision Blizzard} 
	\label{fig:n-day add}
\end{figure*}

In order to calculate the additive abnormal returns we again employ Equation \ref{eq:AAR}. After computing the $AAR_{it}$s for the estimation period and the event periods as discussed in Section \ref{method 1} we draw 3, 5, 7, 9 and 11 one-day abnormal returns from the estimation period $AAR$s. We then calculate the value of $ACAR_i$ for each of these scenarios with the help of Equation \ref{ACAR}. Figure \ref{fig:n-day add} shows the empirical distribution of $ACAR$ for Activision Blizzard. Lastly, to asses the effect of DDoS attack announcement on the stock returns we check the position of $ACAR_i$ for a certain event period in the empirical distribution of $ACAR$ for the same number of days of firm $i$. For example, if we are evaluating the $ACAR$ of Activision Blizzard for event period $[t-1,t+1]$ then we check the position of this $ACAR$ in the 3-day empirical distribution for Activision Blizzard. In this study we consider the 10 percentile scenarios in the left tail to be representative of negative impact and 10 percentile scenarios to the right for positive impact. Hence, if $ACAR_i$ is negative and lies in the bottom 10 percentile of the 5 million scenarios then the impact on the stock returns is considered to be negative. 

\subsubsection{Method 3}
\label{method 3}

In this final method we use a multiplicative model for the estimation of stock returns. The multiplicative estimation model is shown in Equation \ref{multi_model}.

\begin{equation}
\label{multi_model}
(1+r_{it}) = \alpha_i(1+r_{mt})^{\beta_i}\\
\end{equation}

Also, this time we also deviate from the wide spread practice of adding the corresponding single-day returns to compute the cumulative returns. Instead we calculate the exact cumulative returns\footnote{An increase of 10\%, followed by a 10\% decrease implies a total decrease of 1\% according to the multiplicative formula $(1.1)(0.9)=0.99$. The additive approximation yields a change of 0\%, which is an overestimation of 1\%.}.

We linearize Equation \ref{multi_model} as Equation \ref{log model}. The stochastic variable $\epsilon_{it}$ represents the error term with $\E{[\epsilon_{it}]}=0$.

\begin{equation}
\label{log model}
\ln(1+r_{it}) = \widehat{\ln(\alpha_i)}+\hat{\beta_i}\ln(1+r_{mt})+\epsilon_{it}
\end{equation}

\begin{figure*}[h]
	\centering
	\begin{subfigure}[b]{0.475\textwidth}
		\centering
		\includegraphics[width=\textwidth]{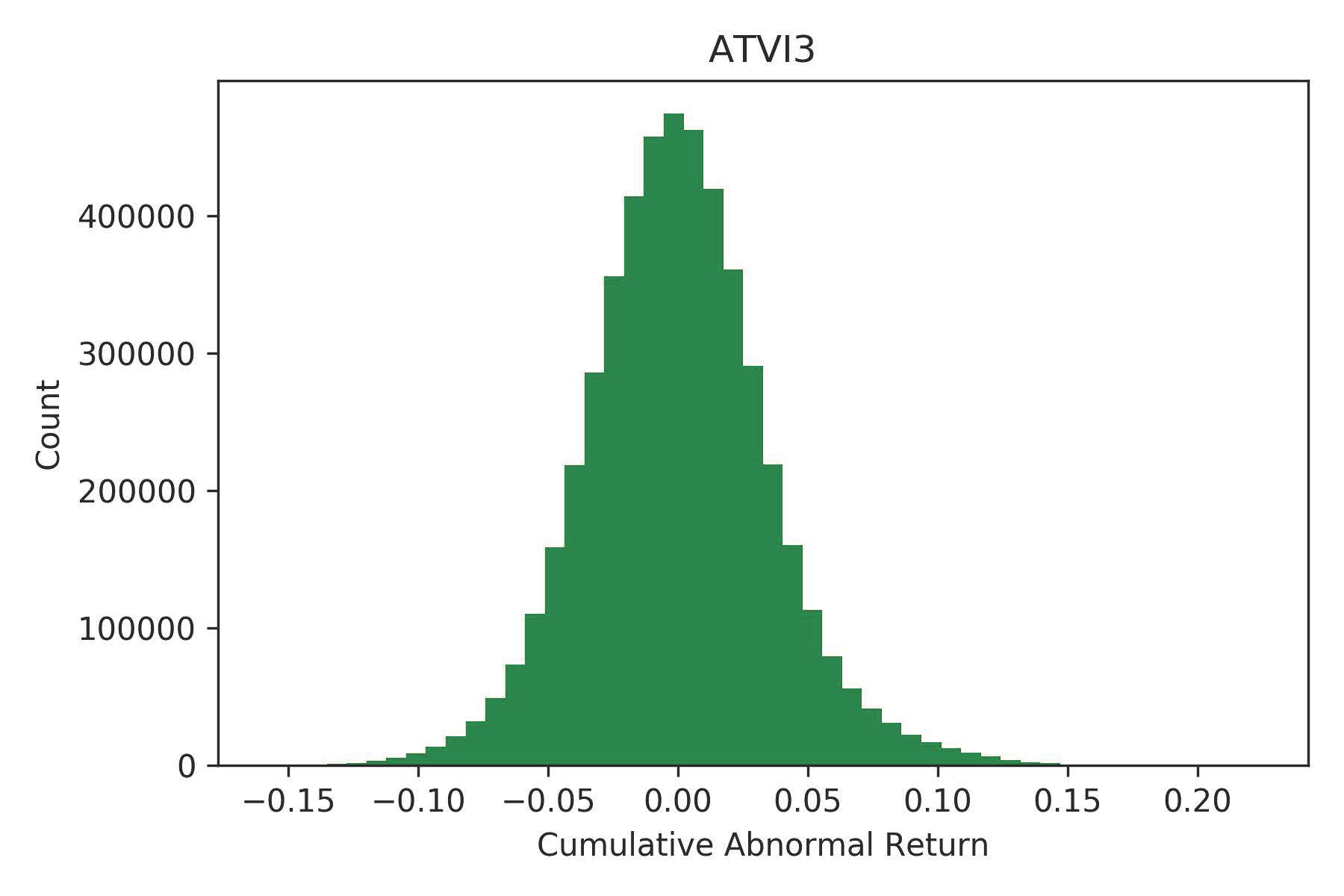}
		\caption[]%
		{{\small 3-Day $CAR_{Activision Blizzard}$}}    
		\label{3-day multi}
	\end{subfigure}
	\hfill
	\begin{subfigure}[b]{0.475\textwidth}  
		\centering 
		\includegraphics[width=\textwidth]{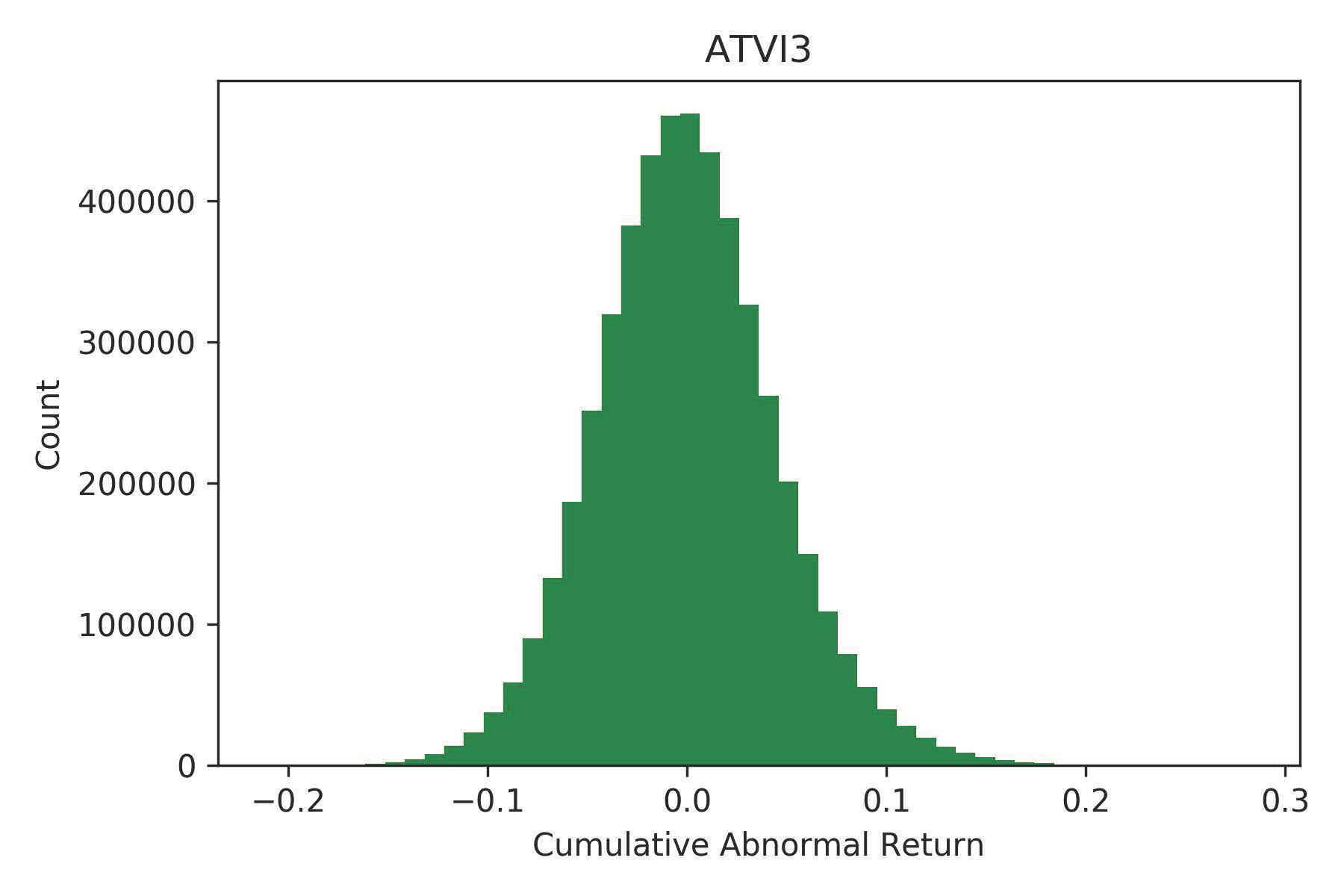}
		\caption[]%
		{{\small 5-Day $CAR_{Activision Blizzard}$}}    
		\label{5-day multi}
	\end{subfigure}
	\vskip\baselineskip
	\begin{subfigure}[b]{0.475\textwidth}   
		\centering 
		\includegraphics[width=\textwidth]{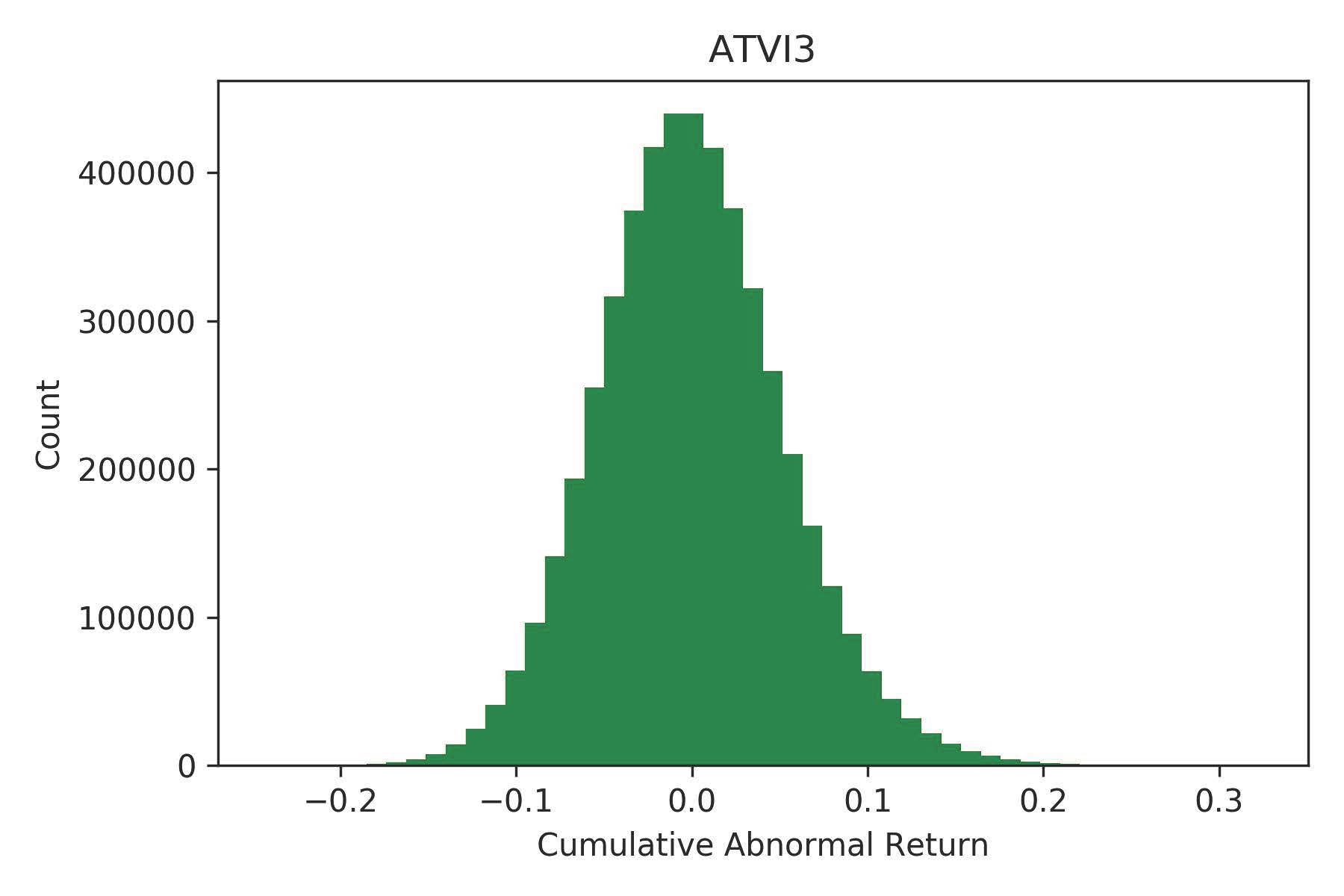}
		\caption[]%
		{{\small 7-Day $CAR_{Activision Blizzard}$}}    
		\label{7-day multi}
	\end{subfigure}
	\quad
	\begin{subfigure}[b]{0.475\textwidth}   
		\centering 
		\includegraphics[width=\textwidth]{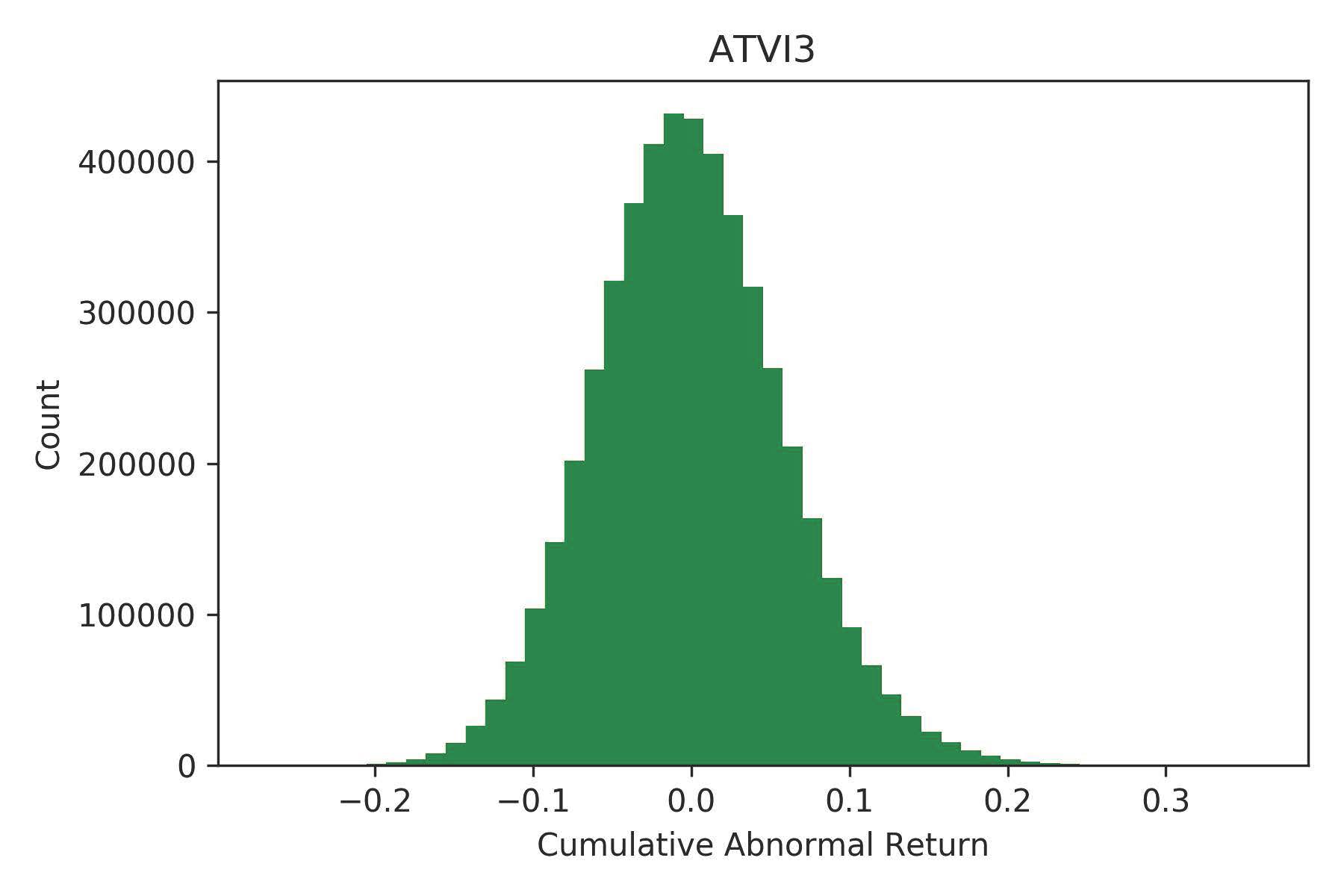}
		\caption[]%
		{{\small 9-Day $CAR_{Activision Blizzard}$}}    
		\label{9-day multi}
	\end{subfigure}
	\caption[]
	{\small Empirical distribution of $CAR$(multiplicative) for Activision Blizzard} 
	\label{fig:n-day multi}
\end{figure*}

After estimating the stock returns we use Equation \ref{transformed} for computing the abnormal returns. As $\widehat{\ln(\alpha_i)}$ is not an unbiased estimator for $\alpha_i$ ($\E{[\hat{\alpha}]}\neq\E{[e^{\widehat{\ln{\alpha}}}]}$), we use Equation \ref{alpha estimator} for estimating $\hat{\alpha}$.

\begin{equation}
\label{transformed}
AR_{it} = \frac{(1+r_{it})}{\hat{\alpha_i}(1+r_{mt})^{\hat{\beta_i}}}-1\\
\end{equation}

\begin{equation}
\label{alpha estimator}
\hat{\alpha_i}=\dfrac{\sum_{t=1}^{T}(1+r_{it})}{\sum_{t=1}^{T}(1+r_{mt})^{\hat{\beta_i}}},
\end{equation}

After computing the $AR_{it}$s for the estimation period and the event periods as discussed in Section \ref{method 1} we draw 3, 5, 7, 9 and 11 one-day abnormal returns from the estimation period $AR$s. As discussed earlier we then calculate the value of $CAR_i$ for each of these scenarios with the help of Equation \ref{CAR}.

\begin{equation}
\label{CAR}
CAR= \prod_{t=N_1}^{N_2}(1+AR_{it})-1
\end{equation}

Figure \ref{fig:n-day multi} shows the empirical distribution of $CAR$ for Activision Blizzard. Lastly, to asses the effect of DDoS attack announcements on the stock returns we check the position of $CAR_i$ for a certain event period in the empirical distribution of $CAR$ for the same number of days of firm $i$. For example, if we are evaluating the $CAR$ of Activision Blizzard for event period $[t-1,t+1]$ then we check the position of this $CAR$ in the 3-day empirical distribution for Activision Blizzard. In this study we consider the 10 percentile scenarios in the left tail to be representative of negative impact and 10 percentile scenarios to the right for positive impact. Hence, if $CAR_i$ is negative and lies in the bottom 10 percentile of the 5 million scenarios then the impact on the stock returns is considered to be negative.

In the next section we discuss the results of our analysis and compare the results.

\section{Results and Discussion}
\label{Results}

%check the dates and argue the results on the basis fo which infrastructure was hit and what was the impact for the customers.

\begin{table}[ht!]
	\centering
	\resizebox{\textwidth}{!}{
		\begin{tabular}{c c | c c c|c c c|c c c|}
			\cmidrule{3-11}
			& & & Method 1 & & & Method 2 & & & Method 3 & \\
			\midrule
			Company Name & Date & +ve periods & -ve periods & No impact & +ve periods & -ve periods & No impact & +ve periods & -ve periods & No impact\\
			\midrule
			Master Card &  2010-12-07 & 2 & 1 & 2 & 2 & 0 & 3 & 2 & 0 & 3\\
			Visa &  2010-12-07 & 2 & 2 & 1 & 2 & 1 & 2 & 2 & 1 & 2\\
			Bank of America &  2010-12-27 & 0 & 3 & 2 & 0 & 3 & 2 & 0 & 3 & 2\\
			Vodafone &  2011-10-04 & 0 & 0 & 5 & 0 & 0 & 5 & 0 & 0 & 5\\
			Vivendi &  2012-01-18 & 0 & 0 & 5 & 0 & 0 & 5 & 0 & 0 & 5\\
			Bursa Malaysia &  2012-02-13 & 0 & 0 & 5 & 0 & 0 & 5 & 0 & 0 & 5\\
			Apple &  2012-05-25 & 0 & 1 & 4 & 0 & 0 & 5 & 0 & 0 & 5\\
			AT\&T &  2012-08-15 & 0 & 0 & 5 & 1 & 0 & 4 & 1 & 0 & 4\\
			Wells Fargo &  2012-12-19 & 0 & 0 & 5 & 0 & 0 & 5 & 0 & 0 & 5\\
			JP Morgan Chase &  2013-03-12 & 0 & 0 & 5 & 3 & 0 & 2 & 3 & 0 & 2\\
			TD Canada Trust &  2013-03-20 & 0 & 0 & 5 & 0 & 1 & 4 & 0 & 1 & 4\\
			American Express &  2013-03-27 & 0 & 0 & 5 & 1 & 0 & 4 & 1 & 0 & 4\\
			ING &  2013-04-08 & 0 & 3 & 2 & 0 & 2 & 3 & 0 & 2 & 3\\
			Linkedin &  2013-06-20 & 0 & 1 & 4 & 0 & 0 & 5 & 0 & 0 & 5\\
			Microsoft &  2013-11-26 & 0 & 0 & 5 & 0 & 0 & 5 & 0 & 0 & 5\\
			RBS &  2013-12-03 & 0 & 0 & 5 & 0 & 0 & 5 & 0 & 0 & 5\\
			Electronic Arts &  2014-01-02 & 0 & 0 & 5 & 0 & 0 & 5 & 0 & 0 & 5\\
			JP Morgan Chase &  2014-01-29 & 0 & 0 & 5 & 0 & 0 & 5 & 0 & 0 & 5\\
			Bank of America &  2014-01-29 & 0 & 0 & 5 & 0 & 0 & 5 & 0 & 0 & 5\\
			Facebook &  2014-02-20 & 0 & 0 & 5 & 0 & 0 & 5 & 0 & 0 & 5\\
			Verizon Communications &  2014-03-21 & 0 & 0 & 5 & 0 & 0 & 5 & 0 & 0 & 5\\
			Activision Blizzard &  2014-03-28 & 1 & 0 & 4 & 2 & 0 & 3 & 2 & 0 & 3\\
			Danske Bank &  2014-07-09 & 0 & 0 & 5 & 0 & 0 & 5 & 0 & 0 & 5\\
			Storebrand &  2014-07-09 & 0 & 0 & 5 & 0 & 0 & 5 & 0 & 0 & 5\\
			Gjensidige Forsikr &  2014-07-09 & 0 & 3 & 2 & 0 & 4 & 1 & 0 & 4 & 1\\
			Sony &  2014-08-22 & 0 & 0 & 5 & 0 & 0 & 5 & 0 & 0 & 5\\
			Amazon &  2014-08-26 & 0 & 0 & 5 & 0 & 0 & 5 & 0 & 0 & 5\\
			Activision Blizzard &  2014-11-13 & 2 & 1 & 2 & 1 & 2 & 2 & 1 & 2 & 2\\
			Sony &  2014-11-25 & 0 & 0 & 5 & 0 & 0 & 5 & 0 & 0 & 5\\
			Rackspace &  2014-12-19 & 0 & 0 & 5 & 0 & 0 & 5 & 0 & 0 & 5\\
			Microsoft &  2014-12-23 & 0 & 0 & 5 & 3 & 0 & 2 & 3 & 0 & 2\\
			Sony &  2014-12-23 & 0 & 0 & 5 & 0 & 0 & 5 & 0 & 0 & 5\\
			Alibaba &  2014-12-24 & 1 & 0 & 4 & 0 & 0 & 5 & 0 & 0 & 5\\
			Nordea Bank &  2015-01-09 & 0 & 3 & 2 & 0 & 3 & 2 & 0 & 3 & 2\\
			Facebook &  2015-01-26 & 0 & 0 & 5 & 0 & 0 & 5 & 0 & 0 & 5\\
			Amazon &  2015-03-13 & 0 & 0 & 5 & 0 & 0 & 5 & 0 & 0 & 5\\
			Electronic Arts &  2015-03-17 & 0 & 4 & 1 & 0 & 1 & 4 & 0 & 1 & 4\\
			Ziggo (Liberty Global) &  2015-08-17 & 2 & 0 & 3 & 4 & 0 & 1 & 4 & 0 & 1\\
			Overstock.com &  2015-09-02 & 0 & 0 & 5 & 0 & 0 & 5 & 0 & 0 & 5\\
			Nissan &  2016-01-12 & 1 & 0 & 4 & 0 & 0 & 5 & 0 & 0 & 5\\
			HSBC &  2016-01-28 & 3 & 0 & 2 & 3 & 0 & 2 & 3 & 0 & 2\\
			Activision Blizzard &  2016-08-02 & 0 & 1 & 4 & 0 & 0 & 5 & 0 & 0 & 5\\
			Electronic Arts &  2016-08-31 & 0 & 1 & 4 & 0 & 0 & 5 & 0 & 0 & 5\\
			StarHub &  2016-10-26 & 0 & 0 & 5 & 2 & 0 & 3 & 2 & 0 & 3\\
			Deutsche Telekom &  2016-11-28 & 0 & 1 & 4 & 0 & 2 & 3 & 0 & 2 & 3\\
			\bottomrule
		\end{tabular}}
		\caption{List of victim companies and summary of results}
		\label{data and summary}
	\end{table}
	
	We now compare the results of our analysis. Table \ref{data and summary} summarizes the outcomes of using the three different methods. The table shows the number of positive and negative \emph{event periods} in each case. A negative event periods imply that the DDoS attack announcement did impact investor decisions. The positive event periods on the stock price actually show that the stock was well performing and the DDoS attack announcement did not have any impact on the stock price. Later in Appendix \ref{Results Table} we present the impact on each firm analyzed in detail.
	
	\begin{table}[h!]
		\centering
		\begin{tabular}{|l||*{3}{c|}}\hline
			\backslashbox{Method 2}{Method 3}
			&\makebox[2em]{+ve}&\makebox[2em]{No}&\makebox[2em]{-ve}\\\hline\hline
			+ve & 24 & 0 & 0 \\\hline
			No  & 0 & 182 & 0 \\\hline
			-ve & 0 & 0 & 19\\\hline
		\end{tabular}
		\caption{Cross-table showing the number of differences between Method 2 and Method 3.} 
		\label{diff:2-3} 
	\end{table}
	
	First we compare the differences in the results when using Method 2 and Method 3. Both methods do not take the assumption of normal distribution for assessing cumulative abnormal returns. However, Method 2 uses an additive model for estimation and Method 3 uses a multiplicative model for the return rate estimation. We find no differences between the results of the two models in the periods analyzed. Hence, we can conclude that the additive model does provide a good estimation for the computation of cumulative abnormal returns.

	Then we look for differences in the results of Method 1 and Method 3. The differences between the models are as follows:
	
	\begin{itemize}
		\item Method 1 uses additive estimation model while Method 2 employs the multiplicative model.
		\item Method 1 computes cumulative abnormal returns by adding the successive abnormal returns where as Method 2 calculates them by using the multiplicative approach (Equation \ref{CAR}).
		\item Finally, Method 3 does not assume the abnormal returns or cumulative abnormal returns to be normally distributed.
	\end{itemize}
	
	\begin{table}[h]
		\centering
		\begin{tabular}{|l||*{3}{c|}}\hline
			\backslashbox{Method 1}{Method 3}
			&\makebox[2em]{+ve}&\makebox[2em]{No}&\makebox[2em]{-ve}\\\hline\hline
			+ve & 11 & 3 & 0 \\\hline
			No  & 13 & 169 & 4 \\\hline
			-ve & 0 & 10 & 15 \\\hline
		\end{tabular}
		\caption{Cross-table showing the number of differences between Method 1 and Method 3.}
		\label{diff:1-3}    
	\end{table}
	
	Table \ref{diff:1-3} summarizes the differences between the two methods. We believe that Method 3 is more accurate, or rather less inaccurate, than Method 1 due to the reduced number of assumptions and approximations in the model. Hence, look at the number of times Method 1 overestimates or underestimates the significance of the results, i.e. gives a significant positive or negative impact when there is no impact or vice-versa. We observe that Method 1 overestimates the significance of the abnormal returns  5.77\% (total 225 periods are considered in this study) of the times and underestimates it 7.55\% of the times. We find these differences to be consistent between Method 1 and Method 2 as well. This suggests that the assumption of normally distributes abnormal returns accounts for these inconsistencies between the results of Method 1 and Method 3 (or Method 2).
	
	\section{Conclusion}
	\label{conclusion}
	
	As an outcome of our study we draw two main conclusions. Firstly, by comparing the various methods of conducting event studies we bring out the risk of overestimating or underestimating the impact of DDoS attack announcements on victim's stock prices. The choice of additive or multiplicative model does not affect the results but the assumption of normally distributed cumulative returns can lead to an incorrect estimation of the impact. Hence, in this study we propose the use of an empirical distribution in order to check the significance of cumulative abnormal returns. Secondly, we also re-emphasize on the results of our previous study \cite{Joosten2017}, and show that all three methods result in a significantly negative event periods on stock price when service to the customers was hampered due to the attack. We reported that the attacks on ING and Nordea bank \cite{ING,Storebrand} resulted in significant negative returns where as Visa and Mastercard \cite{VisaandMastercard} resulted in no damage. Similarly, in case of the attack on Deutsche Telekom that drove nearly 1 million of its customers offline \cite{Deutsche}, we observe a negative impact on the stock price in the 9-day and 11-day period.
	
	%\cleardoublepage
	\printbibliography

	\newpage
	\appendix
	\section{Impact on victim stock prices}
	\label{Results Table}
	\resetmidruleV
	\begin{longtable}{c H c c c c<{\midruleV}}
		\toprule
		Firm & Date & Event Period & Method 1 & Method 2 & Method 3\\
		\midrule

		\endhead
		\noalign{\resetmidruleV}%
		
		&  2012-02-14 & 3-day & No & No & No\\
		&  2012-02-16 & 5-day & No & No & No\\
		Bursa Malaysia &  2012-02-21 & 7-day & No & No & No\\
		&  2012-02-23 & 9-day & No & No & No\\
		&  2012-02-27 & 11-day & No & No & No\\
		&  2012-05-29 & 3-day & -ve & No & No\\
		&  2012-05-31 & 5-day & No & No & No\\
		Apple &  2012-06-04 & 7-day & No & No & No\\
		&  2012-06-06 & 9-day & No & No & No\\
		&  2012-06-08 & 11-day & No & No & No\\
		&  2015-03-16 & 3-day & No & No & No\\
		&  2015-03-18 & 5-day & No & No & No\\
		Amazon &  2015-03-20 & 7-day & No & No & No\\
		&  2015-03-24 & 9-day & No & No & No\\
		&  2015-03-26 & 11-day & No & No & No\\
		&  2014-08-27 & 3-day & No & No & No\\
		&  2014-08-29 & 5-day & No & No & No\\
		Amazon &  2014-09-03 & 7-day & No & No & No\\
		&  2014-09-05 & 9-day & No & No & No\\
		&  2014-09-09 & 11-day & No & No & No\\
		&  2014-11-14 & 3-day & +ve & No & No\\
		&  2014-11-18 & 5-day & +ve & +ve & +ve\\
		Activision Blizzard &  2014-11-20 & 7-day & No & No & No\\
		&  2014-11-24 & 9-day & -ve & -ve & -ve\\
		&  2014-11-26 & 11-day & No & -ve & -ve\\
		&  2014-03-31 & 3-day & No & No & No\\
		&  2014-04-02 & 5-day & No & No & No\\
		Activision Blizzard &  2014-04-04 & 7-day & +ve & +ve & +ve\\
		&  2014-04-08 & 9-day & No & No & No\\
		&  2014-04-10 & 11-day & No & +ve & +ve\\
		&  2016-08-03 & 3-day & -ve & No & No\\
		&  2016-08-05 & 5-day & No & No & No\\
		Activision Blizzard &  2016-08-09 & 7-day & No & No & No\\
		&  2016-08-11 & 9-day & No & No & No\\
		&  2016-08-15 & 11-day & No & No & No\\
		&  2013-03-28 & 3-day & No & No & No\\
		&  2013-04-02 & 5-day & No & No & No\\
		American Express &  2013-04-04 & 7-day & No & No & No\\
		&  2013-04-08 & 9-day & No & +ve & +ve\\
		&  2013-04-10 & 11-day & No & No & No\\
		&  2014-12-26 & 3-day & No & No & No\\
		&  2014-12-30 & 5-day & No & No & No\\
		Alibaba &  2015-01-02 & 7-day & +ve & No & No\\
		&  2015-01-06 & 9-day & No & No & No\\
		&  2015-01-08 & 11-day & No & No & No\\
		&  2010-12-27 & 3-day & No & No & No\\
		&  2010-12-29 & 5-day & No & No & No\\
		Bank of America &  2010-12-31 & 7-day & -ve & -ve & -ve\\
		&  2011-01-04 & 9-day & -ve & -ve & -ve\\
		&  2011-01-06 & 11-day & -ve & -ve & -ve\\
		&  2014-01-30 & 3-day & No & No & No\\
		&  2014-02-03 & 5-day & No & No & No\\
		Bank of America &  2014-02-05 & 7-day & No & No & No\\
		&  2014-02-07 & 9-day & No & No & No\\
		&  2014-02-11 & 11-day & No & No & No\\
		&  2016-10-27 & 3-day & No & No & No\\
		&  2016-10-31 & 5-day & No & No & No\\
		StarHub &  2016-11-02 & 7-day & No & No & No\\
		&  2016-11-04 & 9-day & No & +ve & +ve\\
		&  2016-11-08 & 11-day & No & +ve & +ve\\
		&  2014-07-10 & 3-day & No & No & No\\
		&  2014-07-14 & 5-day & No & No & No\\
		Danske Bank &  2014-07-16 & 7-day & No & No & No\\
		&  2014-07-18 & 9-day & No & No & No\\
		&  2014-07-22 & 11-day & No & No & No\\
		&  2016-11-29 & 3-day & No & No & No\\
		&  2016-12-01 & 5-day & No & No & No\\
		Deutsche Telekom &  2016-12-05 & 7-day & No & No & No\\
		&  2016-12-07 & 9-day & -ve & -ve & -ve\\
		&  2016-12-09 & 11-day & No & -ve & -ve\\
		&  2015-03-18 & 3-day & -ve & No & No\\
		&  2015-03-20 & 5-day & -ve & -ve & -ve\\
		Electronic Arts &  2015-03-24 & 7-day & No & No & No\\
		&  2015-03-26 & 9-day & -ve & No & No\\
		&  2015-03-30 & 11-day & -ve & No & No\\
		&  2014-01-03 & 3-day & No & No & No\\
		&  2014-01-07 & 5-day & No & No & No\\
		Electronic Arts &  2014-01-09 & 7-day & No & No & No\\
		&  2014-01-13 & 9-day & No & No & No\\
		&  2014-01-15 & 11-day & No & No & No\\
		&  2016-09-01 & 3-day & -ve & No & No\\
		&  2016-09-06 & 5-day & No & No & No\\
		Electronic Arts &  2016-09-08 & 7-day & No & No & No\\
		&  2016-09-12 & 9-day & No & No & No\\
		&  2016-09-14 & 11-day & No & No & No\\
		&  2015-01-27 & 3-day & No & No & No\\
		&  2015-01-29 & 5-day & No & No & No\\
		Facebook &  2015-02-02 & 7-day & No & No & No\\
		&  2015-02-04 & 9-day & No & No & No\\
		&  2015-02-06 & 11-day & No & No & No\\
		&  2014-02-21 & 3-day & No & No & No\\
		&  2014-02-25 & 5-day & No & No & No\\
		Facebook &  2014-02-27 & 7-day & No & No & No\\
		&  2014-03-03 & 9-day & No & No & No\\
		&  2014-03-05 & 11-day & No & No & No\\
		&  2014-07-10 & 3-day & No & No & No\\
		&  2014-07-14 & 5-day & -ve & -ve & -ve\\
		Gjensidige Forsikr &  2014-07-16 & 7-day & -ve & -ve & -ve\\
		&  2014-07-18 & 9-day & -ve & -ve & -ve\\
		&  2014-07-22 & 11-day & No & -ve & -ve\\
		&  2016-01-29 & 3-day & No & No & No\\
		&  2016-02-02 & 5-day & +ve & +ve & +ve\\
		Activision Blizzard &  2016-02-04 & 7-day & No & No & No\\
		&  2016-02-08 & 9-day & +ve & +ve & +ve\\
		&  2016-02-10 & 11-day & +ve & +ve & +ve\\
		&  2013-04-09 & 3-day & -ve & -ve & -ve\\
		&  2013-04-11 & 5-day & -ve & No & No\\
		ING &  2013-04-15 & 7-day & -ve & -ve & -ve\\
		&  2013-04-17 & 9-day & No & No & No\\
		&  2013-04-19 & 11-day & No & No & No\\
		&  2014-01-30 & 3-day & No & No & No\\
		&  2014-02-03 & 5-day & No & No & No\\
		JP Morgan Chase &  2014-02-05 & 7-day & No & No & No\\
		&  2014-02-07 & 9-day & No & No & No\\
		&  2014-02-11 & 11-day & No & No & No\\
		&  2013-03-13 & 3-day & No & No & No\\
		&  2013-03-15 & 5-day & No & No & No\\
		JP Morgan Chase &  2013-03-19 & 7-day & No & +ve & +ve\\
		&  2013-03-21 & 9-day & No & +ve & +ve\\
		&  2013-03-25 & 11-day & No & +ve & +ve\\
		&  2015-08-18 & 3-day & No & No & No\\
		&  2015-08-20 & 5-day & +ve & +ve & +ve\\
		Ziggo (Liberty Global) &  2015-08-24 & 7-day & +ve & +ve & +ve\\
		&  2015-08-26 & 9-day & No & +ve & +ve\\
		&  2015-08-28 & 11-day & No & +ve & +ve\\
		&  2013-06-21 & 3-day & No & No & No\\
		&  2013-06-25 & 5-day & No & No & No\\
		Linkedin &  2013-06-27 & 7-day & No & No & No\\
		&  2013-07-01 & 9-day & No & No & No\\
		&  2013-07-03 & 11-day & -ve & No & No\\
		&  2010-12-08 & 3-day & No & No & No\\
		&  2010-12-10 & 5-day & -ve & No & No\\
		Master Card &  2010-12-14 & 7-day & No & No & No\\
		&  2010-12-16 & 9-day & +ve & +ve & +ve\\
		&  2010-12-20 & 11-day & +ve & +ve & +ve\\
		&  2013-11-27 & 3-day & No & No & No\\
		&  2013-12-02 & 5-day & No & No & No\\
		Microsoft &  2013-12-04 & 7-day & No & No & No\\
		&  2013-12-06 & 9-day & No & No & No\\
		&  2013-12-10 & 11-day & No & No & No\\
		&  2014-12-24 & 3-day & No & No & No\\
		&  2014-12-29 & 5-day & No & +ve & +ve\\
		Microsoft &  2014-12-31 & 7-day & No & +ve & +ve\\
		&  2015-01-05 & 9-day & No & +ve & +ve\\
		&  2015-01-07 & 11-day & No & No & No\\
		&  2015-01-12 & 3-day & No & No & No\\
		&  2015-01-14 & 5-day & No & No & No\\
		Nordea Bank &  2015-01-16 & 7-day & -ve & -ve & -ve\\
		&  2015-01-21 & 9-day & -ve & -ve & -ve\\
		&  2015-01-23 & 11-day & -ve & -ve & -ve\\
		&  2016-01-13 & 3-day & No & No & No\\
		&  2016-01-15 & 5-day & No & No & No\\
		Nissan &  2016-01-20 & 7-day & +ve & No & No\\
		&  2016-01-22 & 9-day & No & No & No\\
		&  2016-01-26 & 11-day & No & No & No\\
		&  2015-09-03 & 3-day & No & No & No\\
		&  2015-09-08 & 5-day & No & No & No\\
		Overstock.com &  2015-09-10 & 7-day & No & No & No\\
		&  2015-09-14 & 9-day & No & No & No\\
		&  2015-09-16 & 11-day & No & No & No\\
		&  2014-12-22 & 3-day & No & No & No\\
		&  2014-12-24 & 5-day & No & No & No\\
		Rackspace &  2014-12-29 & 7-day & No & No & No\\
		&  2014-12-31 & 9-day & No & No & No\\
		&  2015-01-05 & 11-day & No & No & No\\
		&  2013-12-04 & 3-day & No & No & No\\
		&  2013-12-06 & 5-day & No & No & No\\
		RBS &  2013-12-10 & 7-day & No & No & No\\
		&  2013-12-12 & 9-day & No & No & No\\
		&  2013-12-16 & 11-day & No & No & No\\
		&  2014-12-24 & 3-day & No & No & No\\
		&  2014-12-29 & 5-day & No & No & No\\
		Sony &  2014-12-31 & 7-day & No & No & No\\
		&  2015-01-05 & 9-day & No & No & No\\
		&  2015-01-07 & 11-day & No & No & No\\
		&  2014-08-25 & 3-day & No & No & No\\
		&  2014-08-27 & 5-day & No & No & No\\
		Sony &  2014-08-29 & 7-day & No & No & No\\
		&  2014-09-03 & 9-day & No & No & No\\
		&  2014-09-05 & 11-day & No & No & No\\
		&  2014-11-26 & 3-day & No & No & No\\
		&  2014-12-01 & 5-day & No & No & No\\
		Sony &  2014-12-03 & 7-day & No & No & No\\
		&  2014-12-05 & 9-day & No & No & No\\
		&  2014-12-09 & 11-day & No & No & No\\
		&  2014-07-10 & 3-day & No & No & No\\
		&  2014-07-14 & 5-day & No & No & No\\
		Storebrand &  2014-07-16 & 7-day & No & No & No\\
		&  2014-07-18 & 9-day & No & No & No\\
		&  2014-07-22 & 11-day & No & No & No\\
		&  2012-08-16 & 3-day & No & No & No\\
		&  2012-08-20 & 5-day & No & +ve & +ve\\
		AT\&T &  2012-08-22 & 7-day & No & No & No\\
		&  2012-08-24 & 9-day & No & No & No\\
		&  2012-08-28 & 11-day & No & No & No\\
		&  2013-03-21 & 3-day & No & No & No\\
		&  2013-03-25 & 5-day & No & No & No\\
		TD Canada Trust &  2013-03-27 & 7-day & No & No & No\\
		&  2013-04-01 & 9-day & No & -ve & -ve\\
		&  2013-04-03 & 11-day & No & No & No\\
		&  2010-12-08 & 3-day & -ve & No & No\\
		&  2010-12-10 & 5-day & -ve & -ve & -ve\\
		Visa &  2010-12-14 & 7-day & No & No & No\\
		&  2010-12-16 & 9-day & +ve & +ve & +ve\\
		&  2010-12-20 & 11-day & +ve & +ve & +ve\\
		&  2012-01-19 & 3-day & No & No & No\\
		&  2012-01-23 & 5-day & No & No & No\\
		Vivendi &  2012-01-25 & 7-day & No & No & No\\
		&  2012-01-27 & 9-day & No & No & No\\
		&  2012-01-31 & 11-day & No & No & No\\
		&  2011-10-05 & 3-day & No & No & No\\
		&  2011-10-07 & 5-day & No & No & No\\
		Vodafone &  2011-10-11 & 7-day & No & No & No\\
		&  2011-10-13 & 9-day & No & No & No\\
		&  2011-10-17 & 11-day & No & No & No\\
		&  2014-03-24 & 3-day & No & No & No\\
		&  2014-03-26 & 5-day & No & No & No\\
		Verizon Communications &  2014-03-28 & 7-day & No & No & No\\
		&  2014-04-01 & 9-day & No & No & No\\
		&  2014-04-03 & 11-day & No & No & No\\
		&  2012-12-20 & 3-day & No & No & No\\
		&  2012-12-24 & 5-day & No & No & No\\
		Wells Fargo &  2012-12-27 & 7-day & No & No & No\\
		&  2012-12-31 & 9-day & No & No & No\\
		&  2013-01-03 & 11-day & No & No & No\\
	\end{longtable}
	*The multiple events related to the same firm are sorted date wise.

%------------------------------------------------------------------------------
\end{document}

% EOF